# Band Engineering of Carbon Nitride Monolayers by N-type, P-type, and Isoelectronic Doping for Photocatalytic Applications

Meysam Makaremi[1], Sean Grixti[1], Keith T. Butler[2], Geoffrey A. Ozin[3] and Chandra Veer Singh[*,1,4]

[1]Department of Materials Science and Engineering, University of Toronto, 184 College Street, Suite 140, Toronto, ON M5S 3E4, Canada

[2]Centre for Sustainable Chemical Technologies and Department of Chemistry, University of Bath, Bath, BA2 7AY, United Kingdom

[3]Department of Chemistry, Solar Fuels Research Cluster, University of Toronto, 80 St. George Street, Toronto, ON M5S 3H6, Canada

[4]Department of Mechanical and Industrial Engineering, University of Toronto, 5 King's College Road, Toronto, ON M5S 3G8, Canada

ABSTRACT Since hydrogen fuel involves the highest energy density among all fuels, production of this gas through the solar water splitting approach has been suggested as a green remedy for greenhouse environmental issues due to extensive consumption of fossil fuels. Low dimensional materials possessing a large surface-to-volume ratio can be a promising candidate to be used for the photocatalytic approach. Here, we used extensive first principles calculations to investigate the application of newly fabricated members of two-dimensional carbon nitrides including *tg*-$C_3N_4$, *hg*-$C_3N_4$, $C_2N$, and $C_3N$ for water splitting. Band engineering via n-type, p-type, and isoelectronic doping agents such as B, N, P, Si, and Ge was demonstrated for tuning the electronic structure; optimizing solar absorption and band alignment for photocatalysis. Pristine *tg*-$C_3N_4$, *hg*-$C_3N_4$, and $C_2N$ crystals involve bandgaps of 3.190 eV, 2.772 eV, and 2.465 eV, respectively, which are not proper for water splitting. Among the dopants, Si and Ge dopants can narrow the band gap of carbon nitrides about 0.5 – 1.0 eV, and also increase their optical absorption in the visible spectrum. This study presents the potential for doping with isoelectronic elements to greatly improve the photocatalytic characteristics of carbon nitride nanostructures.

Keywords: carbon nitride, water splitting, density functional theory, band structure, band edge position, solar absorption spectrum.

## 1. Introduction

Rapidly growing human population and extensive fossil fuel energy consumption leading to environmental issues have motivated scientists to device new techniques to extract and store reliable green-energy from renewable natural sources such as wind, hydro, biomass and sunlight. Specifically sustainable energy systems and applications involving production of hydrogen fuel from water have been the research focus during the recent years.[1–4] For the first time in 1972, water was split into $H_2$ and $O_2$ through a solar electrochemical approach and the breaking of water molecule emerged as a clean and reliable remedy for future energy issues.[5–7] In solar water splitting, $H_2$, which is a clean fuel and contains the highest energy density (~ 142 MJ kg$^{-1}$) among fuels, is the final product.[8–10]

Progress in the green energy industry can greatly benefit from contemporary advances in the development of low dimensional nanostructures such as nanoparticles, nanotubes, and two dimensional (2D) materials which demonstrate fascinating properties for optoelectronic, catalytic, and energy storage/conversion applications stemming from their large surface-to-volume ratios.[11,12] There had been a great enthusiasm to extract 2D graphene from the bulk structure of graphite for decades until finally Novoselov *et al.* mechanically exfoliated graphite and isolated the carbon monolayer in 2004.[13] Physical properties of graphene, including fascinating thermal, optical, mechanical and electronic characteristics, generated a flurry of research activity to uncover other 2D materials.[14–16]

In the last decade, a broad spectrum of 2D materials has been predicted, fabricated and characterized.[17–19] These materials may involve only single elements; such as graphene, germanene and stanene;[20] or they might contain binary/multiplinary structures;

such as boron nitride (BN)[21,22], transition metal oxides (TMOs)[23,24], transition metal dichalcogenides (TMDs)[25–28], and metal nitrides/carbides/carbonitrides (MXenes)[29,30]. Two approaches including top-down and bottom-up techniques are employed to fabricate the monolayers.[31–33] The former generates the nanosheet from the bulk structure by the means of physical exfoliation, while the latter synthesizes the monolayer by linking the unit blocks via chemical reaction.[34] Recently carbon nitride nanosheets (2D-CN) including $C_3N_4$[35,36], $C_2N$[37], and $C_3N$[38] have been synthesized through bottom-up procedures.

2D-CN nanostructures show outstanding optical, thermal, mechanical and electronic properties due to their strong atomic networks composed of C and N atoms which have comparable atomic sizes and contains four and five valence electrons, respectively, forming consistent covalent configurations.[39–41] $C_3N_4$ can exist in different configurations including cubic phase, semi-cubic phase, $\alpha$-phase, $\beta$-phase, and graphitic (g) phase in two forms (*hg*-$C_3N_4$ and *tg*-$C_3N_4$) among which g-phases are known to be the most stable phases with the non-metallic nature including energy gaps of 2.7 and 3.1 eV.[42–44] Nitrogenated holey graphene is another carbon nitride nanosheet with a stoichiometric formula of $C_2N$ which contains an evenly distributed network of N and hole sites, which makes it an excellent candidate as a nanofilter for shape and size selective adsorption of different ions, atoms and molecules.[37,45] Lately, 2D polyaniline with one N and three C atoms per unitcell ($C_3N$) have been fabricated and it is found to have amazing optical, thermal, mechanical, electronic, and magnetic properties.[46–48] Additional degrees of freedom are provided by the ability to dope these materials, facilitating the engineering of band structures to tailor the system for a given application. This wide compositional and structural, however, is daunting to explore exhaustively by synthesis and characterization.

Computational modelling allows for understanding and the development of general design principles. Whilst there has been limited theoretical studies, an in depth survey of trends across the range of different carbon nitride nanosheets, with various doping regimes, is still lacking.[49–52] The recent successful fabrication of different carbon nitride nanosheets including *hg*-$C_3N_4$, *tg*-$C_3N_4$, $C_2N$, and $C_3N$ with fascinating semiconducting behaviors prompted us to consider the possible application of these materials for the solar water splitting by tuning the bandgap. [35–38] We carried out extensive density functional theory (DFT) simulations to tune the bandgap through n-type, p-type, and isoelectronic doping with different elements consisting of B, N, P, Si, and Ge dopants. We used different electronic structure calculations including structural relaxation, adsorption energy, electronic density of states, band energy alignment, and absorbance spectrum analyses.

## 2. Computational Details

Perdew–Burke–Ernzerhof (PBE)[53] and Heyd-Scuseria-Ernzerhof (HSE06)[54] density functional theory (DFT) techniques implemented in the Vienna *Ab-initio* Simulation Package (VASP)[55], were employed via generalized gradient approximation (GGA) and projector augmented-wave (PAW) potentials[56]. A kinetic energy cutoff of 500 eV, electronic self-consistency of $1 \times 10^{-6}$ eV and ionic relaxation convergence of $1 \times 10^{-3}$ eV/Å were applied. Also a Grimme dispersion correction technique, DFT-D2[57] was considered to modify van der Waals energy calculations. Monkhorst-Pack grids of $15 \times 15 \times 1$ and $6 \times 6 \times 1$ were used for PBE and HSE06 calculations; respectively, and the tetrahedron scheme with Blöchl corrections was employed to integrate the Brillouin zone.

The optical response of 2D structures was evaluated by complex dielectric function calculations.[58] The function composed of the real ($\epsilon^1$) and imaginary ($\epsilon^2$) parts which can be determined by,

$$\epsilon^1(\omega) = 1 + \frac{2}{\pi} P \int \frac{\epsilon^2(\omega')\omega'}{\omega'^2 - \omega^2} d\omega', \tag{1}$$

$$\epsilon^2(\omega) = \frac{4\pi^2 e^2}{m^2 \omega^2} \sum_{c,v} \int \frac{|P_{c,v}(k)|^2}{\nabla \omega_{c,v}(k)} dC_k, \tag{2}$$

here, $C_k$ and $P$ are the surface-energy constant and the principle part of the $\epsilon^1$ integral, respectively. $P_{c,v}$ and $\omega_{c,v}$ are dipole and energy difference matrix element between conduction ($c$) and valence ($v$) states, respectively. The absorption coefficient $\alpha(\omega)$ can be described as,

$$\alpha(\omega) = \sqrt{2}\omega \left\{ \sqrt{\epsilon^1(\omega)^2 + \epsilon^2(\omega)^2} - \epsilon^1(\omega) \right\}^{\frac{1}{2}}, \tag{3}$$

The unit cells of each carbon nitride nanosheet used are described as follows: the unit cell of C$_2$N has 12 carbon atoms and 6 nitrogen atoms, and it can be characterized as two interconnected benzene rings via a pyrazine ring. The C$_3$N unit cell has 6 carbon atoms and 2 nitrogen atoms, and has a honeycomb structure, similar to graphene. The *t*g-C$_3$N$_4$ unit cell has 3 carbon atoms and 4 nitrogen atoms, and is characterized as interconnected triazine molecules bridged by nitrogen atoms. The *hg*-C$_3$N$_4$ unit cell has 6 carbon atoms and 8 nitrogen atoms, and is characterized as an array of interconnected heptazine molecules bridged by nitrogen atoms.

### 3. Results and Discussion

The relaxed atomic structures of pristine carbon nitride nanosheets including C$_2$N, C$_3$N, *t*g-C$_3$N$_4$, and *hg*-C$_3$N$_4$ can be seen in Figure 1. The lattice parameters of the PBE relaxed $C_2N$ unit cell were determined to be a=8.263Å and γ=60°, which is in good agreement with previous simulations and experimental data.[37,52,59] The lattice parameters of the PBE relaxed $C_3N$ unit cell were determined to be a=4.861Å and γ=120°, which is also in good agreement with previous results.[38,60] The lattice parameters of the PBE relaxed *tg*-$C_3N_4$ unit cell were determined to be a=4.783Å and γ=120°, which is in good agreement with previous results.[61] The lattice parameters of the PBE relaxed *hg*-$C_3N_4$ unit cell were determined to be a=7.133Å and γ=120°, which is also in good agreement with previous results.[35,61]

Furthermore, The electron localization function (ELF)[62] of different CN structures is depicted in Figure 1. The normalized ELF contour involves a spectrum ranging from values of 0 to 1, in which 0 (blue) and 1 (red) present the lack and abundance of electron localization, respectively. If the ELF contour is located at the center of a bond, it shows covalent bonding, while if the pronounced localization contour is located on one side of the bond (on one of the atoms), it illustrates ionic bonding.[63] Each CN structure involves two different bonds including C-C and C-N bonds. The electron localization at the middle of both bonds for each lattice illustrates covalent bonding between C-C and C-N atoms; Moreover, CN crystals contain two kinds of N atoms with respect to the number of bonded atoms, including two or three C atoms. There is a charge localization on the former type of N atoms, showing these bonds have stronger ionic properties and weaker covalent nature compared to the latter type.

Doping each carbon nitride on the carbon site was attempted with B, N, P, Si, and Ge as dopants. The relaxed structure of the successfully doped systems can be seen in Figure 2 and Figure S1 structural characteristics of the doped carbon nitride materials can be seen in Tables 1, 2, S1, and S2. We note that the doping of Ge into the carbon site of $C_3N$ and $tg$-$C_3N_4$ causes the structural collapse for $C_3N$, and the formation of a new phase completely different from the initial $tg$-$C_3N_4$ structure, respectively, and so their properties are not reported. A trend of increasing structural deformation with increasing dopant atomic size is noted, as expected, with distortions in atomic structure and unit cell lattice parameters being present. From the successfully doped systems electronic structure calculations were performed to determine their applicability for photocatalytic water splitting applications.

From Tables 1, 2 and 3 the band gap of each pristine and doped carbon nitride systems is reported. The band gaps obtained from using the PBE functional for $C_2N$, $C_3N$, $hg$-$C_3N_4$ and $tg$-$C_3N_4$ were 1.660 eV, 0.386 eV, 1.197 eV, and 1.574 eV, respectively. Since the PBE functional is known to underestimate band gaps hybrid functional calculations with the HSE-06 functional were performed on each of the relaxed structures. The band gaps obtained from using the HSE-06 functional for $C_2N$, $C_3N$, $hg$-$C_3N_4$, and $tg$-$C_3N_4$ were 2.465 eV, 1.049 eV, 2.772 eV and 3.190 eV, respectively. The band gaps obtained from the HSE-06 functional are more than double the value of the band gaps obtained from the PBE functional. The band gaps obtained from the HSE-06 functional match literature values very well, as seen in Table 3, with previous reports of 2.47 eV, 1.042 eV, 2.72 eV, and 3.1 eV for $C_2N$, $C_3N$, $hg$-$C_3N_4$, and $tg$-$C_3N_4$, respectively.[35,52,60,64]

In Figure 3 the total density of states and projected orbital density of states of each pristine system is shown. It is noted there is significant hybridization between the carbon 2p states and the nitrogen 2p states in all systems. The $C_p$-$N_p$ hybridization is the main bonding source in all pristine systems, originating from the combination of $sp^2$-$sp^2$ σ-bonding and p-p π-bonding. Significant hybridization between the carbon p-states, nitrogen p-states, and the dopant p-states is observed in all doped systems as well. When doped with aliovalent dopants, such as B, N and P, all carbon nitride nanosheets become metallic. This is due to the significant p- and n-doping from the doping species. B p-dopes each system significantly, pushing the Fermi level into the valence band making each system metallic, as seen in Figure S2. On the other hand, N and P both n-dope each system significantly, pushing the fermi level into the conduction band, making each system metallic, as seen in Figure S2.

When doped with isoelectronic dopants the semiconducting character of each system is conserved, therefore band gap engineering is possible. When $C_2N$ is doped with Si and Ge the band gap narrows to 1.754 eV and 1.810 eV, respectively, from the 2.465 eV band gap of the pristine system. When $C_3N$ is doped with Si the band gap narrows from 1.042 eV to 0.331 eV. When $tg$-$C_3N_4$ is doped with Si the band gap narrows to 2.209 eV from 3.190 eV. Band gap narrowing is also observed when doping $hg$-$C_3N_4$ in the corner site with Si and Ge, and in bay site with Ge, with the band gap narrowing from 2.772 eV to 2.385 eV, 2.508 eV, and 2.691 eV, respectively. This band gap narrowing is consistent with previous investigations into isoelectronic doping of graphene materials.[52,65]

The band gap narrowing that occurs with the isoelectronic dopants can be explained by a decrease bond strength and hybridization due to smaller orbital overlap. The larger dopant atom increases the bond distance and increases the ionicity of the bonding character, and therefore, decreases the orbital interaction with the smaller nitrogen and carbon atoms.[66] Since the valence band maximum (VBM) has a bonding character and the conduction band minimum (CBM) has an anti-bonding character, the decreased hybridization results in an effective upwards shift of the VBM and downward shift of the CBM. However, it is noted that contrary to group IV band gap trends, Si provides greater band narrowing compared to Ge.[67–69] This can be explained by the added bonding interaction from the 3d orbital of Ge hybridizing with the sp$^2$ orbitals of the neighboring atoms, as observed from the projected density of states in Figure 4.

Band widening occurs when Si is doped into the bay site of *hg*-$C_3N_4$, increasing the band gap from 2.772 eV to 2.886 eV. This deviation from the trend can be explained by the fact the bond distance does not significantly increase from pristine upon the addition of Si and greater orbital overlap of the Si 3p orbitals with the neighboring 2p states is observed. This effect is not seen when Ge is doped into the bay site of *hg*-$C_3N_4$, even though the bond distance of doesn't increase significantly from pristine. Band widening does not occur when Ge is doped into the bay site as the iconicity of the bonding character is greater, compared to Si, as observed by a lower Ge-4p density of states in Figure 4h.

The splitting of water involves two redox half reactions,

$$\text{Reduction: } 2H^+(aq) + 2e^- \rightarrow H_2(g), \tag{4}$$

$$\text{Oxidation: } 2H_2O(l) \rightarrow O_2(g) + 4H^+(aq) + 4e^-. \tag{5}$$

at pH = 0, the reduction potential of H$^+$/H$_2$ and the oxidation potential of O$_2$/H$_2$O are -4.44 eV and -5.67 eV, respectively, therefore the minimal theoretical energy gap for a material to be applied for the solar water splitting process needs to be 1.23 eV.[70–72] As a consequence, among all of the 2D carbon nitride structures studied in this work, pristine $C_2N$, *tg*-$C_3N_4$, and *hg*-$C_3N_4$ crystals and their counterparts doped with Si and Ge warrant further investigation for the photocatalytic application.

The configuration of band edges is a key factor determining the applicability of a semiconductor for photocatalysis.[9] To be applied for solar water splitting, a material must possess a VBM less negative than the $H^+/H_2$ reduction potential and a CBM energy level more negative than the $O_2/H_2O$ oxidation potential. Semiconducting 2D carbon nitride crystals including pristine and doped $C_2N$, $tg$-$C_3N_4$, and $hg$-$C_3N_4$ structures are illustrated in Figure 5, and compared with the reduction $H^+/H_2$ and oxidation $O_2/H_2O$ energy levels of water splitting at two acidic (pH of 0) and basic (pH of 14) conditions.

Assessing the applicability of material for catalytic processes requires us to place the energies of active carriers in the material in relation to the redox energies of the reactions desired.[73] We use alignment of the valence band maximum (for holes) and the conduction band minimum (for electrons) to place the carrier energies in our systems with respect to the redox potentials required for water splitting, this is depicted in Figure 5. The redox potentials of water at the extremes of pH are presented. We find that under acidic conditions only $C_2N$ and $C_{2-x}Ge_xN$ are capable of reducing hydrogen, while most materials have a VBM capable of oxidising water for oxygen evolution. In the extreme basic conditions, many of the systems have suitable band edge positions for both reactions, with the exception of $tg$-$C_{3-x}Si_xN_4$ (very shallow VBM) and $C_2N$, $C_{2-x}Si_xN$ and $C_{2-x}Ge_xN$ (very deep CBMs). The trends show another powerful demonstration that a mixture of the structural and compositional engineering provide a powerful tool for tuning the absolute band edge positions, as well as the band gap.[74]

Figure 6 shows how changing the parent structure can affect the region of optical absorption of these materials. The $C_2N$ parent structure shows the strongest optical absorption in the visible light region of the electromagnetic spectrum. Doping with Si and Ge result in lower absorption onsets, commensurate with reduced bandgaps, although the strength of absorption is slightly diminished upon doping, this should be considered when fully assessing the application of doping to improve performance. The *tg*-$C_3N_4$ structure has lower optical absorption of visible light, but again doping with Si leads to increased absorption in the visible range. Likewise, hg-$C_3N_4$ has weaker absorption of visible light than $C_2N$. In *hg*-$C_3N_4$ the effect of doping is particularly pronounced, with Si doping leading to a much earlier optical absorption onset and strong activity in the visible range, demonstrating the importance of composition as well as structure for the realisation of photocatalytic carbon nitride based monolayer materials.

## 4. Conclusions

Extensive HSE and PBE DFT simulations were carried out to engineer band structure properties of newly fabricated 2D carbon nitrides including *tg*-$C_3N_4$, *hg*-$C_3N_4$, $C_2N$, and $C_3N$ by N-type, P-type, and isoelectronic doping agents such as B, N, P, Si, and Ge for the water splitting technique. We used structural stability, total and projected electronic density of states, absorbance spectrum and band edge alignment analyses to probe the carbon nitride structures.

While pristine and doped $C_3N$ crystals contain zero or small band gaps (less than 1.23 eV) which are not suitable for photocatalytic water splitting, bare *tg*-$C_3N_4$, *hg*-$C_3N_4$, and $C_2N$ structures and the ones doped by isoelectronic Si and Ge agents show proper semiconducting properties. Specifically tuning the band structures with isoelectronic agents highly improve the band edge positions and visible absorbance spectra of the newly fabricated 2D carbon nitride structures.

This study shows that our doping technique can be applied to tune the bandgap of 2D carbon nitride nanostructures for photocatalytic water splitting, and we plan to study the effect of various band engineering methods; such as the structural aggregation and functionalization. We hope our study will shed light on developing and designing new photocatalytic low dimensional materials to harvest hydrogen from water by the green solar water splitting approach, and making the renewable method more feasible to both meet the growing energy needs and to reduce greenhouse gas emissions.

## ASSOCIATED CONTENT

**Supporting Information**

The Supporting Information is available.

Two figures illustrating the metallic doped carbon nitride structures and the projected density of states (PDOS) of metallic doped carbon nitrides, and two tables listing the Bond distances of pristine and doped carbon nitrides.

## AUTHOR INFORMATION

**Corresponding Author**

* chandraveer.singh@utoronto.ca

## ACKNOWLEDGMENT

Authors gratefully acknowledge their financial support in parts by Natural Sciences and Engineering Council of Canada (NSERC), University of Toronto, Connaught Global Challenge Award, and Hart Professorship. The computations were carried out through Compute Canada facilities, particularly SciNet and Calcul-Quebec. SciNet is funded by the Canada Foundation for Innovation, NSERC, the Government of Ontario, Fed Dev Ontario, and the University of Toronto, and we gratefully acknowledge the continued support of these supercomputing facilities.

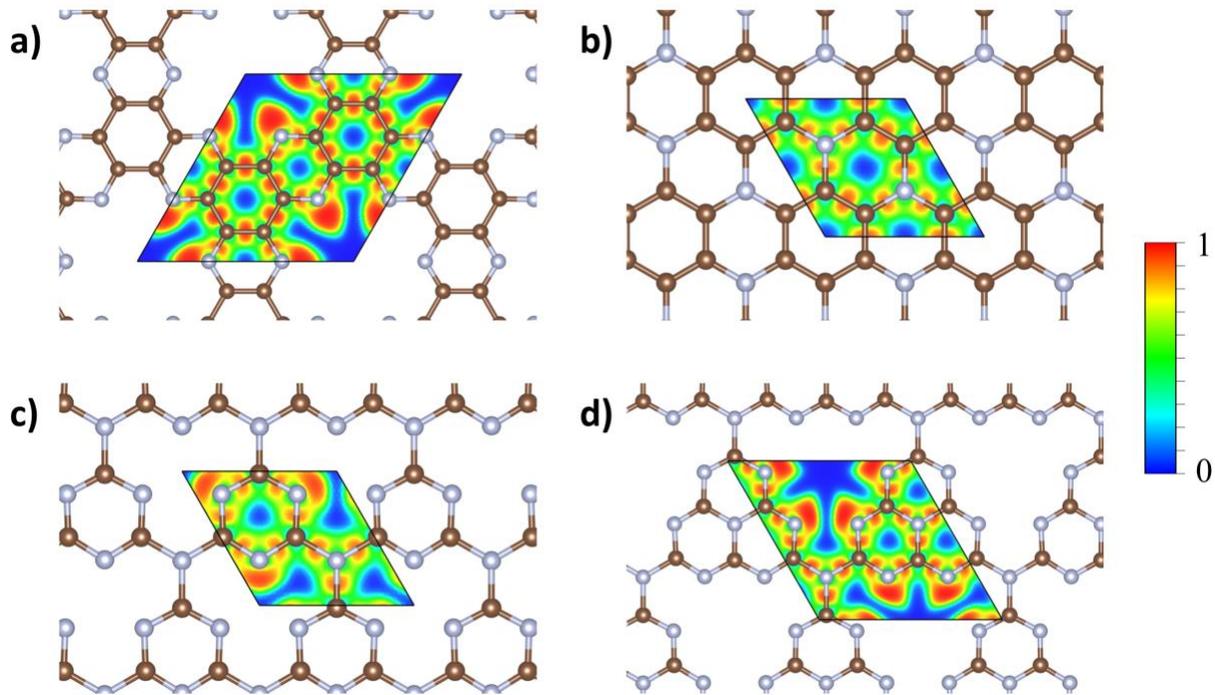

**Figure 1.** Structures of carbon nitrides: a) $C_2N$ b) $C_3N$ c) *tg*-$C_3N_4$ and d) *hg*-$C_3N_4$. The contours illustrate electron localization function (ELF), which has a value between 0 and 1, where 1 corresponds to the perfect localization. With the brown atoms representing carbon and the blue atoms representing nitrogen.

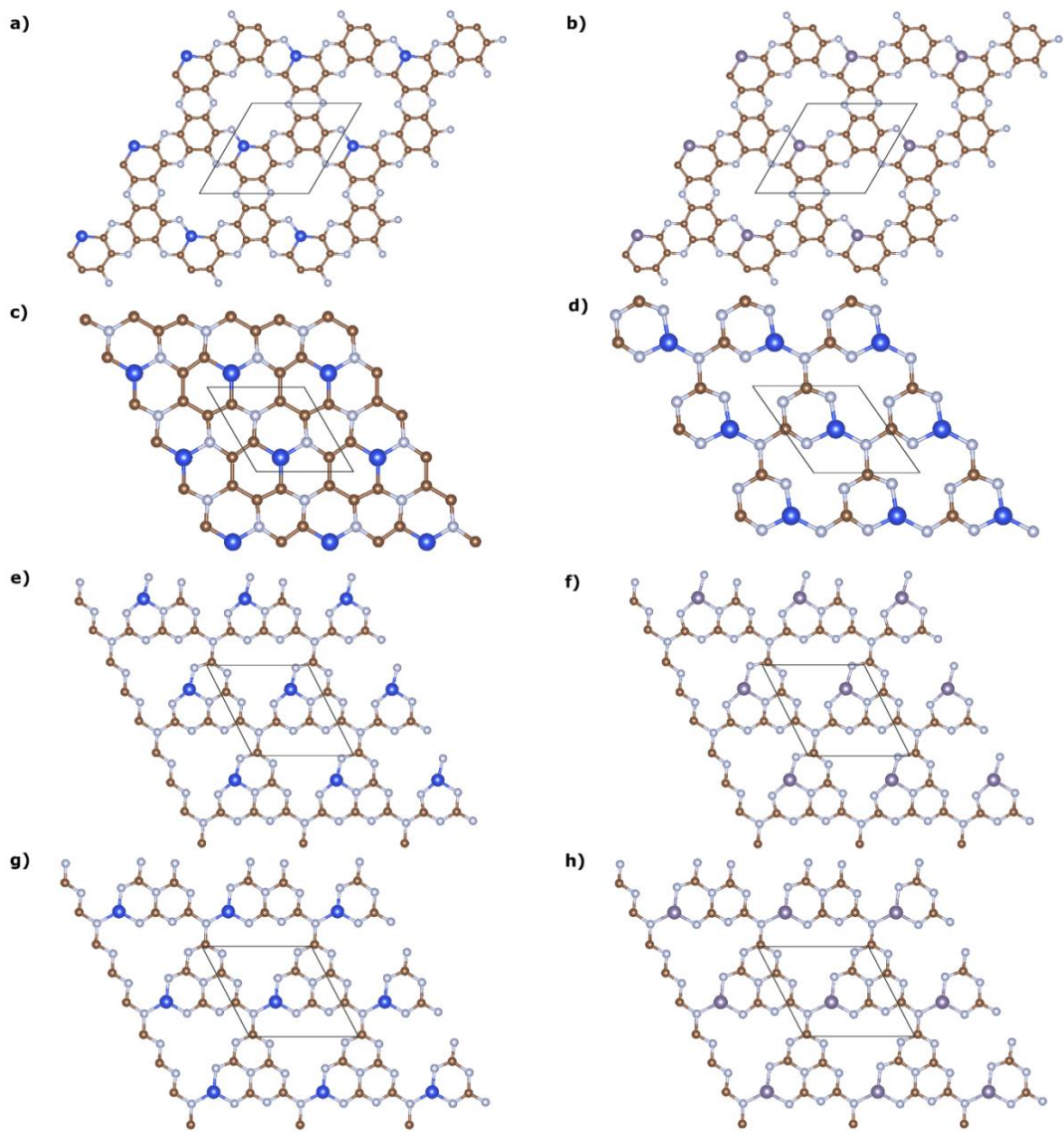

**Figure 2.** Structures of the doped semiconducting carbon nitrides: a) $C_{2-x}Si_xN$, b) $C_{2-x}Ge_xN$, c) $C_{3-x}Si_xN$, d) *tg*-$C_{3-x}Si_xN_4$, e) *hg*-$C_{3-x}Si_xN_4$ *corner site*, f) *hg*-$C_{3-x}Ge_xN_4$ *corner site*, g) *hg*-$C_{3-x}Si_xN_4$ *bay site*, and h) *hg*-$C_{3-x}Ge_xN_4$ *bay site*. Brown atoms represent carbon, light blue atoms represent nitrogen, royal blue atoms represent silicon, and purple represents germanium.

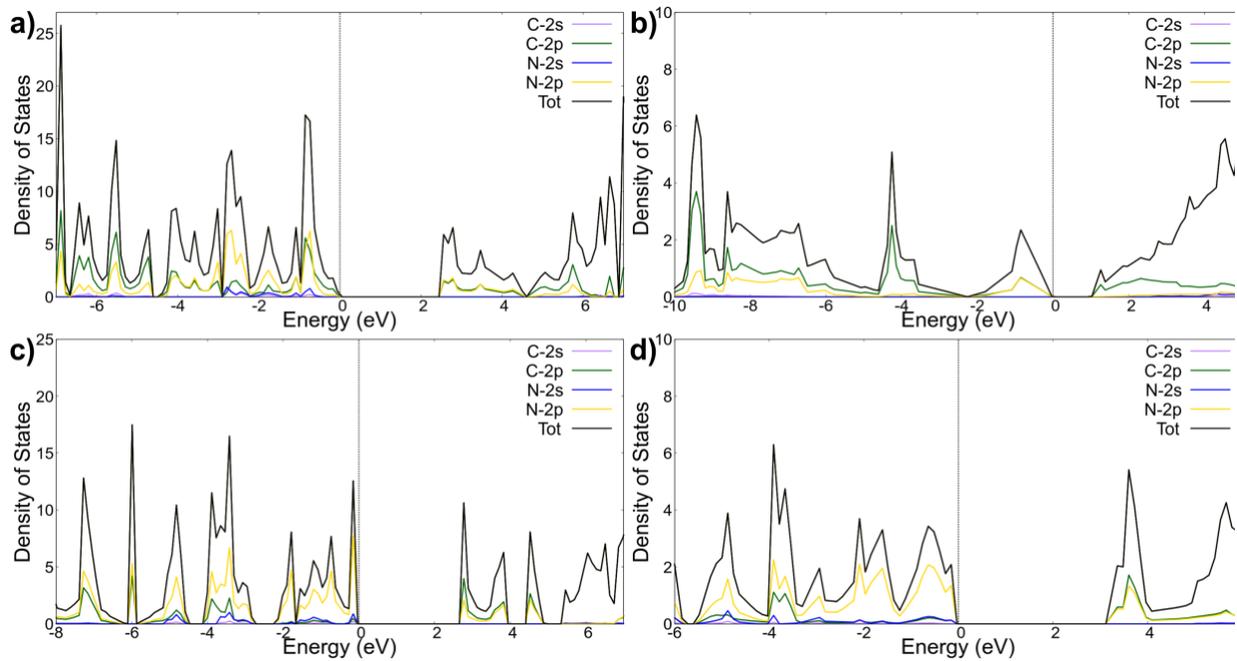

**Figure 3.** Projected density of states (PDOS) of the pristine carbon nitrides: a) $C_2N$, b) $C_3N$, c) *hg*-$C_3N_4$, and d) *tg*-$C_3N_4$

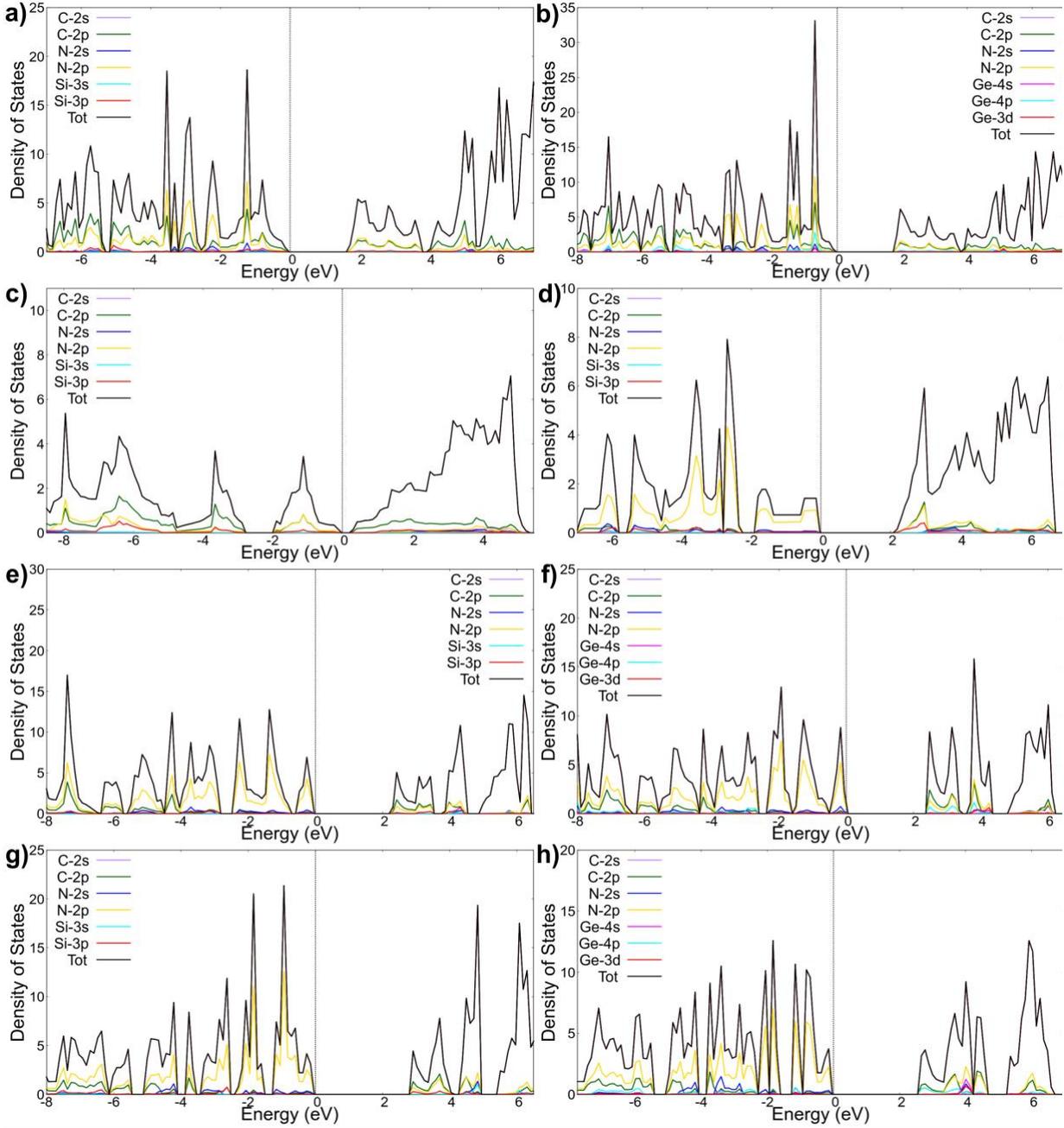

**Figure 4.** Projected density of states (PDOS) of the semiconducting doped carbon nitrides: a) $C_{2-x}Si_xN$, b) $C_{2-x}Ge_xN$, c) $C_{3-x}Si_xN$, d) $tg$-$C_{3-x}Si_xN4$, e) $hg$-$C_{3-x}Si_xN_4$ in the *corner site*, f) $hg$-$C_{3-x}Ge_xN_4$ in the *corner site*, g) $hg$-$C_{3-x}Si_xN_4$ in the *bay site*, h) $hg$-$C_{3-x}Ge_xN_4$ in the *bay site*.

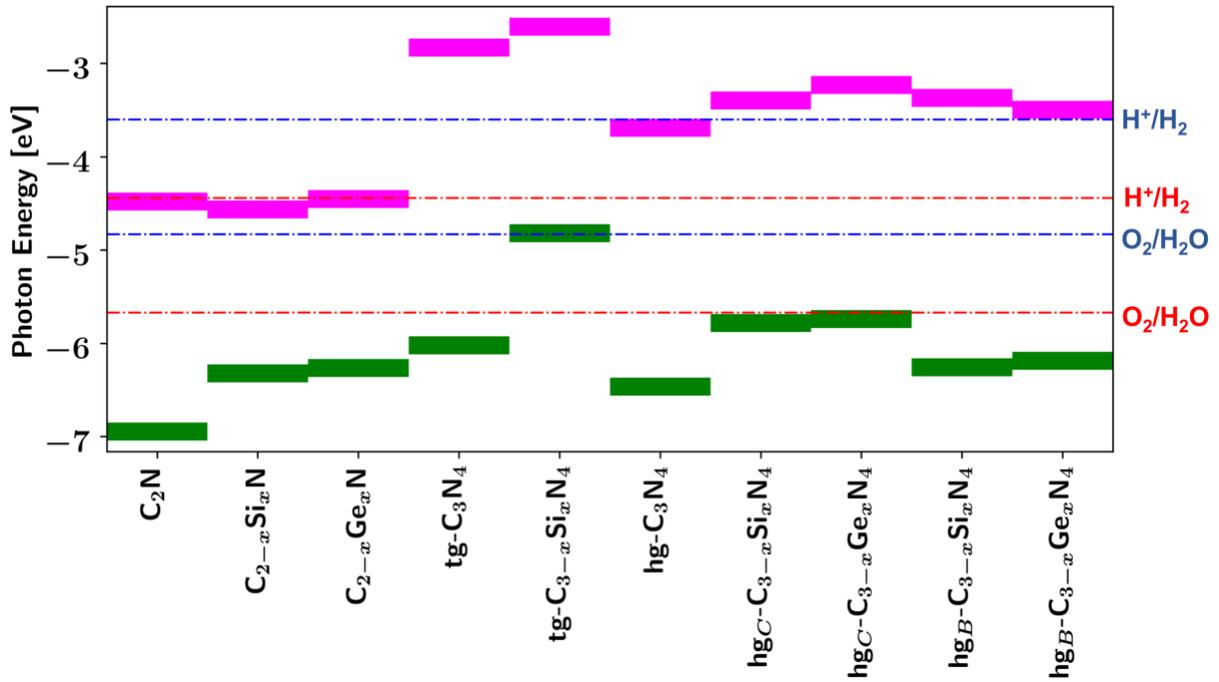

**Figure 5.** Valence (green) and conduction (magenta) band edge positions of pristine and doped carbon nitrides. The energies of the water splitting half reactions at pH 0 (red) and pH 14 (blue) are provided by the dashed lines. A conduction band above $H^+/H_2$ can drive this half reaction; a valence band below $O_2/H_2O$ can drive this half reaction.

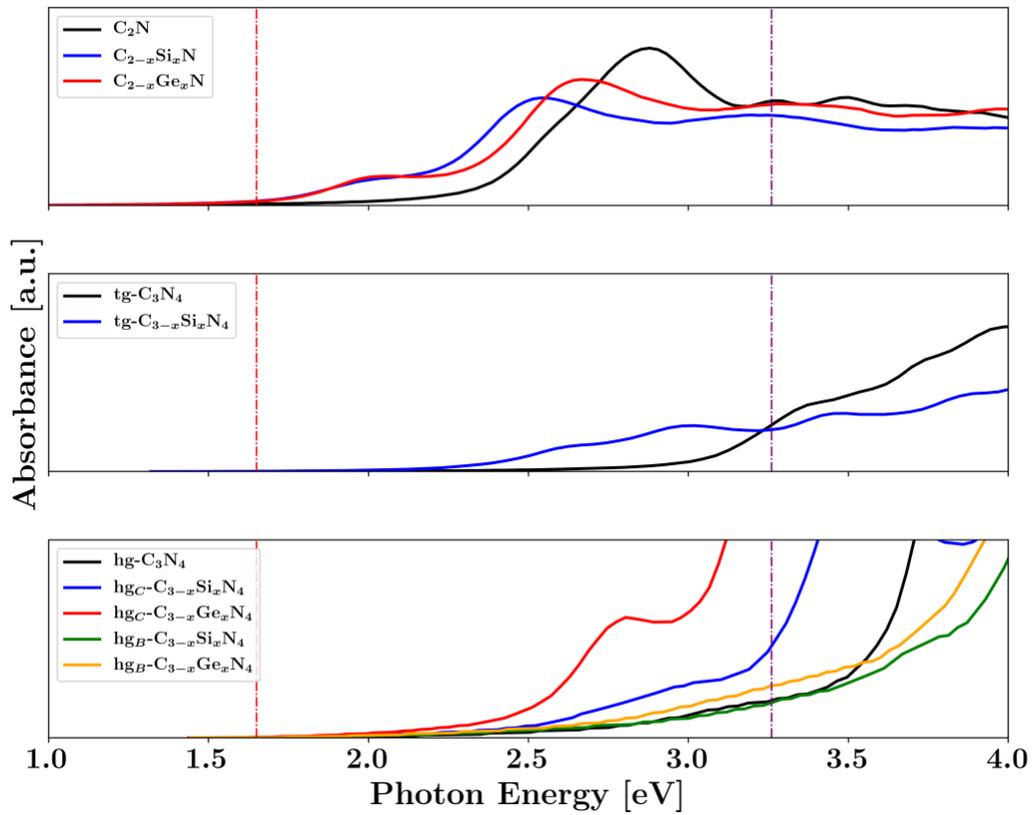

**Figure 6.** Absorption energy spectrum with respect to photon energy for pristine and doped $C_2N$, tg-$C_3N_4$, and hg-$C_3N_4$ structures. The spectra were calculated by using PBE functional corrected by a rigid energy shift considered from HSE06 calculations. Dashed red and purple lines indicate the visible electromagnetic spectrum.

**Table 1:** Structural and electronic characteristics of pristine and doped $C_2N$ and $C_3N$.

| $C_2N$ | | | | | | |
|---|---|---|---|---|---|---|
| Lattice Parameter | Pristine | B | N | P | Si | Ge |
| a (Å) | 8.326 | 8.544 | 8.325 | 8.703 | 8.841 | 8.965 |
| b (Å) | 8.326 | 8.357 | 8.302 | 8.415 | 8.418 | 8.433 |
| γ (°) | 60 | 59.73 | 59.32 | 59.88 | 59.83 | 59.77 |
| | | | | | | |
| Band Gap (eV) | | | | | | |
| PBE | 1.660 | Metallic | Metallic | Metallic | 1.158 | 1.103 |
| HSE-06 | 2.465 | Metallic | Metallic | Metallic | 1.754 | 1.810 |
| $C_3N$ | | | | | | |
| Lattice Parameter | Pristine | B | N | P | Si | |
| a (Å) | 4.861 | 4.927 | 4.831 | 5.139 | 5.193 | |
| b (Å) | 4.861 | 4.948 | 4.831 | 5.156 | 5.232 | |
| γ (°) | 120.00 | 120.00 | 120.00 | 120.11 | 120.25 | |
| | | | | | | |
| Band Gap (eV) | | | | | | |
| PBE | 0.386 | Metallic | Metallic | Metallic | Metallic | |
| HSE-06 | 1.049 | Metallic | Metallic | Metallic | 0.331 | |

**Table 2:** Structural and electronic characteristics of pristine and doped *tg*-C$_3$N$_4$ and *hg*-C$_3$N$_4$.

| *tg*-C$_3$N$_4$ | | | | | | |
|---|---|---|---|---|---|---|
| Lattice Parameter | Pristine | B | N | P | Si | |
| a (Å) | 4.783 | 4.961 | 4.732 | 5.203 | 5.338 | |
| b (Å) | 4.783 | 4.961 | 4.732 | 5.203 | 5.342 | |
| γ (°) | 120 | 124.76 | 119.08 | 123.91 | 125.25 | |
| Band Gap (eV) | | | | | | |
| PBE | 1.574 | Metallic | Metallic | Metallic | 0.890 | |
| HSE-06 | 3.190 | Metallic | Metallic | Metallic | 2.209 | |
| *hg*-C$_3$N$_4$ | | | | | | |
| *Corner Site* | | | | | | |
| Lattice Parameter | Pristine | B | N | P | Si | Ge |
| a (Å) | 7.134 | 7.094 | 7.127 | 7.250 | 7.288 | 7.400 |
| b (Å) | 7.133 | 7.096 | 7.126 | 7.251 | 7.290 | 7.398 |
| γ (°) | 120.00 | 118.12 | 120.44 | 117.71 | 116.80 | 116.78 |
| Band Gap (eV) | | | | | | |
| PBE | 1.197 | Metallic | Metallic | Metallic | 0.913 | 1.070 |
| HSE-06 | 2.772 | Metallic | Metallic | Metallic | 2.385 | 2.508 |
| *Bay Site* | | | | | | |
| Lattice Parameter | Pristine | B | N | P | Si | Ge |
| a (Å) | 7.134 | 7.252 | 7.088 | 7.509 | 7.626 | 7.803 |
| b (Å) | 7.133 | 7.057 | 7.160 | 7.123 | 7.108 | 7.101 |
| γ (°) | 120.00 | 119.12 | 120.35 | 118.30 | 117.69 | 117.33 |
| Band Gap (eV) | | | | | | |
| PBE | 1.197 | Metallic | Metallic | Metallic | 1.389 | 1.248 |
| HSE-06 | 2.772 | Metallic | Metallic | Metallic | 2.886 | 2.691 |

**Table 3:** Band gaps of the pristine carbon nitride nanosheets.

| Band Gap (eV) | $C_2N$ | $C_3N$ | $tg$-$C_3N_4$ | $hg$-$C_3N_4$ |
|---|---|---|---|---|
| Current work (HSE-06) | 2.465 | 1.049 | 3.190 | 2.772 |
| Literature | 2.47[48] | 1.042[56] | 3.1[59] | 2.72[33] |